\documentclass[10pt]{iopart}
\usepackage{graphicx}
\usepackage{amssymb,amsthm,bm} %% ams packages
\usepackage{cfkc}  %% style for type-setting algorithms

\def\CC{\mathbb{C}}
\def\CP{\mathbb{CP}}
\def\SU{\mathbf{SU}}
\def\RR{\mathbb{R}}
\def\H{\mathcal{H}}
\def\vc{\vec{c}}
\def\vt{\vec{\theta}}
\def\vp{\vec{\phi}}
\def\ket#1{|#1\rangle}
\def\bra#1{\langle #1|}
\def\ip#1#2{\langle #1 | #2 \rangle}
\def\up{{\uparrow}}
\def\dn{{\downarrow}}

\begin{document}
\title[Bang-Bang Control Design for Quantum State Transfer]{Bang-Bang Control
       Design for Quantum State Transfer based on Hyperspherical
       Coordinates and Optimal Time-energy Control}

\author{Weiwei Zhou$^1$, S.~G. Schirmer$^2$, Ming Zhang$^1$, Hong-Yi Dai$^3$}

\address{$^1$Department of Automatic Control, College of Mechatronics and
Automation,  National University of Defense Technology,  Changsha
410073,  People's Republicof China}
\address{$^2$Department of Applied Mathematics and Theoretical Physics,
            University of Cambridge, Cambridge, CB3 0WA, UK}
\address{$^3$Department of Physics, College of Science, National University of Defense
            Technology, Changsha  410073, People's Republic of China}
\ead{zwwann@nudt.edu.cn, sgs29@cam.ac.uk, zhangming@nudt.edu.cn,
hydai@nudt.edu.cn}

\begin{abstract}
We present a constructive control scheme for solving quantum state
engineering problems based on a parametrization of the state vector in
terms of complex hyperspherical coordinates.  Unlike many control
schemes based on factorization of unitary operators, the scheme gives
explicit expressions for all generalized Euler angles in terms of the
hyperspherical coordinates of the initial and final states.  The
factorization, when applicable, has added benefits that phase rotations
can be combined and performed concurrently.  The control procedure can
be realized using simple bang-bang or square-wave-function
controls. Optimal time-energy control is considered to find the optimal
control amplitude.  The extension of the scheme to implement arbitrary
unitary operators is also discussed.
\end{abstract}

\noindent\textbf{Keywords:}
quantum systems, Bang-Bang control, geometric parametrization,
controllability, optimal control
%\end{keyword}

\section{Introduction}

Control of phenomena governed by the laws of quantum mechanics is
increasingly recognized as a crucial task and prerequiste to realizing
promising new technologies based on quantum effects from the use of
photonic reagents in chemistry~\cite{Rabitz2000} to quantum metrology
and quantum information processing~\cite{Nielsen2000} to mention only a
few examples.  From early beginnings in the 1980s (see e.g.,
\cite{Huang1983,Ong1984,Clark1985,Blaquiere1987,Pierce1988}), there has
been considerable recent progress in both theory and experiment of
quantum control~\cite{Daless2007,Wiseman2010}.  Among the core tasks for
quantum control are quantum state and operator engineering.  In the
former case the main objective is to prepare the system in a desired
state, which is usually a pure state $\ket{\psi_f}$ represented by a
unit vector in a Hilbert space $\H$ associated with the system.  The
task can take various forms, from state transfer, i.e., steering the
system from a known initial state $\ket{\psi_0}$ to the target
state~\cite{qse1,qse2,qse3,qse4,qse5}, to purification or state
reduction, i.e., preparation of a desired pure state starting with a
mixed or unknown initial state, usually involving some form of feedback
from measurements of an observable~\cite{fb1,fb2,fb3}, to protecting or
stabilizing a desired state in the presence of environment noise or
disturbance \cite{st1,st2,st3}.  Often the goal is the preparation of a
non-classical state such as a Greenberger-Horne-Zeilinger (GHZ) state
\cite{GHZ}, or a maximally entangled Bell state~\cite{Bell}.  Operator
engineering usually involves engineering the dynamical evolution to
realize a particular unitary operator~\cite{uod1,uod2,uod3,uod4,uod5},
and plays a crucial role in the implementation of quantum gates in the
context for quantum information processing.

Some of the tasks mentioned above such as purification or stabilization
generally require measurements and feedback~\cite{fb1,fb2,fb3}, or
possibly coherent feedback~\cite{cfb1,cfb2}, and some proposed control
strategies actively take advantage of environmental effects~\cite{emc1,
emc2,emc3} or even backaction effects of measurements and
feedback~\cite{st1,st2}.  Perhaps the majority of control strategies for
quantum dynamics, at least to date, however, rely on coherent open-loop
control, i.e., manipulation of the dynamics via coherent interaction of
the system with external fields or potentials, the type of control that
we will be considered in this article.  The main reason for foregoing
measurements and feedback is to avoid the disturbance of the system that
results from the backaction of measurement and feedback on the system,
leading to complex non-unitary dynamics and decoherence.  The challenge
of open-loop control is to design external fields or potentials acting
as controls off-line based on a model of the system.  The main
strategies for control design in this context are based either on
geometric ideas, or more formally Lie group decompositions, as in
\cite{uod1,uod2,uod3,uod4,uod5}, so-called model-based feedback design
\cite{Lya1,Lya2,Lya3,Lya4,Lya5,Lya6}, or formulating the problem as an
optimal control problem and using optimization techniques (see
\cite{oct} and references therein, \cite{oct2,oct3}).  The latter
approach has been used successfully to find solutions for many different
types of control problems --- from control of vibrational modes via
ultrashort laser pulses~\cite{oct1}, to control of nuclear spin systems
\cite{oct2,oct3}, to control of spatially distributed systems in
\cite{oct4}, to implemention unitary operators~\cite{oct5,oct6} and
encoded logic gates~\cite{oct7}, to optimizing state transfer in spin
networks~\cite{oct8}, to the creation of various types of
entanglement~\cite{GHZ,Bell}.  Optimal control is important and holds
considerable promise of enabling robust control of complex, imperfect
systems with limited control~\cite{oct5,oct6,oct7}.  Often it requires
control with complex temporal and spectral profiles, however, which may
be difficult to implement for certain systems, e.g., in solid-state
quantum dot systems controlled by voltages applied to gate electrodes,
where it may be difficult to implement complicated time-varying voltage
profiles.

For these reasons constructive control schemes that require only simple
pulses such as approximately piecewise-constant functions (bang-bang
controls) remain useful as an alternative, which can often be optimized
to mitigate limitations of the system or the control to some extent, as
in the case of optimized Euler angles to compensate for non-orthogonal
rotation axes \cite{euler1,euler2}, for example.  This is the type of
control considered here.  Specifically, we consider state-transfer tasks in
which initial and finial states are given.  We show that parametrization
of the initial and target states in hyperspherical coordinates~\cite{gb}
yields a simple constructive control scheme for state-transfer tasks
that requires no complex calculations of the control parameters, i.e.,
all control parameters are given in terms of simple functions of the
initial and final state coordinates.  The scheme has some additional
advantages over alternative geometric schemes, e.g., based on
decomposition into Givens rotations~\cite{uod1} in that many operations
can be performed either sequentially or in parallel, reducing the time
required to implement the control schemes.  We introduce a parameter
$\lambda$ which represents the ratio of costs of time and energy, and
further explore the trade-off between time and energy optimal control
using time-energy performance index $J = \int^{t_{f}}_{0}[\lambda+E(t)]
dt$, where $E(t)$ is energy cost of Bang-Bang control at $t$, $t_{f}$ is
terminal time. It is shown that the product of the terminal time
$t^{*}_{f}$ and the energy cost $E^{*}$ for optimal bounded or unbounded
piecewise constant controls only depends on the geometric parameters of
the initial and target states and is independent of $\lambda$ but
$\lambda$ determines the optimal field strength of the controls,
$L_*=\sqrt{\lambda}$.  The scheme can be generalized to implement
arbitrary unitary operators, and we again find that the resulting
decomposition has some advantages in that many operations commute and
can be performed in parallel.

\section{Pure-state Transfer by Bang-Bang Control}

Pure-states $\ket{\psi}$ of a quantum system defined on a complex
Hilbert space $\H$ with $\dim\H=N<\infty$ can be represented by complex
vectors $\vec{c}\in\CC^N$ by choosing a suitable basis
$\{\ket{n}\}_{n=1}^N$ for $\H$,
\begin{equation}
 \label{3-1} \ket{\psi} = \sum_{n=1}^{N}{c_n}\ket{n}.
\end{equation}
The modulus squared $|c_n|^2$ of the coordinates can be interpreted
in terms of probabilities provided $\vc$ is a unit vector.  For most
applications the global phase of the state is irrelevant, i.e. we
can further identify $\ket{\psi}\sim e^{i\phi}\ket{\psi}$.  Given
these considerations, physically distinguishable pure states can be
uniquely identified with elements in the complex projective space
$\CP^{N-1}=S^{2N-1}/S_1$, and we can uniquely represent pure states
by unit vectors in $\CC^N$ if we fix the complex phase of one
coordinate.

Pure-state transfer is the task of transforming a given pure quantum
state $\ket{\psi^{(0)}}$ to a desired pure quantum state
$\ket{\psi^{(s)}}$ and is one of the most fundamental tasks in control
of quantum systems.  Many of the control strategies mentioned in the
introduction have been applied to this problem, including constructive
control schemes based on the Lie group decomposition.  Indeed, it is
quite straightforward to see how to solve the state transfer problem for
an $N$-level system in principle, if we are able to implement unitary
gates on a sequence of connected two-level subspaces.  Assume, e.g.,
that $\SU(2)$ operations can be implemented on the subspaces spanned by
$\{\ket{1},\ket{2}\}$, $\{\ket{2},\ket{3}\}$, \ldots,
$\{\ket{N-1},\ket{N}\}$.  We can decompose any unitary operator in
$\SU(N)$ into a sequence of $\SU(2)$ rotations on these two-dimensional
(2D) subspaces.  Each of these can be further decomposed into a sequence
of three rotations about two orthogonal axes using the Euler
decomposition. It therefore suffices if we can implement rotations about
two fixed orthogonal axes on each of the 2D subspaces.  Applied to the
problem of quantum state transfer, it is not difficult to see that we
can transform any complex unit vector $\vec{c}^{(0)}$ into any other
complex unit vector $\vec{c}^{(s)}$ by a sequence of $N-1$ rotations on
the 2D subspaces defined above
\begin{equation}
  \vec{c}^{(s)} = U^{(N-1,N)}\ldots U^{(2,3)}U^{(1,2)} \vec{c}^{(0)}
\end{equation}
where $U^{(n,n+1)}$ indicates a complex rotation on the subspace spanned
by $\{\ket{n},\ket{n+1}\}$.  Decomposing each $U^{(n,n+1)}$ further into
three rotations about two fixed orthogonal axes, $U_1^{(n,n+1)}(\alpha)$
and $U_2^{(n,n+1)}(\beta)$, by suitable angles $\gamma_k$,
\begin{equation}
  U^{(n,n+1)} = U_1^{(n,n+1)}(\gamma_3)
                U_2^{(n,n+1)}(\gamma_2)
                U_1^{(n,n+1)}(\gamma_1),
\end{equation}
shows that in general $3(N-1)$ such rotations are required to transform
a given initial state to a target state using a sequence of elementary
unitary transformations,
\begin{multline}
  \label{eq:decomp}
  \vec{c}^{(s)} =
  U_1^{(N-1,N)}(\gamma_{3N-3})
  U_2^{(N-1,N)}(\gamma_{3N-4})
  U_1^{(N-1,N)}(\gamma_{3N-5})
  \times \ldots \\
  \times U_1^{(1,2)}(\gamma_{3})
  U_2^{(1,2)}(\gamma_{2})
  U_1^{(1,2)}(\gamma_{1}) \vec{c}^{(0)}.
\end{multline}
It is easy to see how to transform pure states in principle, but it is
not obvious how to derive the correct rotation angles $\gamma_k$ in the
sequence, which is what matters in practice.  Although it is possible to
constructively compute the $\gamma_k$, the dependence of $\gamma_k$ on
the state vectors $\vec{c}^{(0)}$ and $\vec{c}^{(s)}$ is complicated.

\section{Bang-Bang Control Scheme based on Hyperspherical Parametrization}

In this section we discuss how to obtain explicit expressions for the
rotation angles $\gamma_k$ and show that it can be easily solved by
parameterizing the initial and target states in terms of complex
hyperspherical coordinates.

\subsection{Complex hyperspherical coordinates }

Any complex unit vector $\vec{c}$ can be parametrized in terms of
complex hyperspherical coordinates $(\vt,\vp)$,
\begin{equation}
\label{3-2}
 \begin{pmatrix} c_1\\ c_2\\ \vdots \\ c_{N-1} \\ c_N \end{pmatrix}
 = e^{i\phi_0}
 \left( \begin{array}{l}
  \cos\theta_{1}\\
  e^{i\phi_{1}}\sin\theta_{1}\cos\theta_{2}\\
  \vdots \\
  e^{i\phi_{N-2}}\sin\theta_{1}\ldots \sin\theta_{N-2}\cos\theta_{N-1}\\
  e^{i\phi_{N-1}}\sin\theta_{1}\ldots \sin\theta_{N-1}
 \end{array}\right)
\end{equation}
where $\vt$ and $\vp$ are vectors in $\RR^{N-1}$ with $0\le\theta_n
\le\tfrac{\pi}{2}$ and $-\pi\le\phi_n\le \pi$, and $e^{i\phi_0}$ is
a global phase factor, which is usually negligible.  Thus, assuming
normalization and neglecting global phases, any pure state is
uniquely determined by its complex hyperspherical coordinates
$(\vt,\vp)$ which can be  calculated easily by
Algorithm~\ref{algo1}.

Although there are many equivalent parameterizations of pure state
vectors, the beauty of complex hyperspherical coordinates is that we
can easily give an explicit constructive bang-bang control scheme
for state transfer $\ket{\psi^{(0)}} \mapsto \ket{\psi^{(s)}}$ such
that all control pulses are determined directly by the coordinates
of the initial and final states
$(\vt^{(0)},\vp^{(0)},\vt^{(s)},\vp^{(s)})$.

\begin{KCalgorithm}
\KCin{$c$}{complex vector/pure state}
\KCout{$\theta,\phi$}{hyper-spherical coordinates}
\KCcode{HyperCoord}{Compute complex hyperspherical
 coordinates}{29}{\columnwidth}{
   \KClet{$N$}{\KCid{length}$(c)$}
   \KClet{$c$}{$c/$\KCid{norm}$(c)$}
   \KClet{$c$}{$\exp(-i$ * \KCid{angle}$(c_1))*c$}
   \KClet{$\phi$}{\KCid{angle}$(c_{2:N})$}
   \KClet{$a$}{\KCid{abs}$(c)$}
   \KClet{$\theta_1$}{$\arccos(a_1)$}
   \KClet{$s_1$}{$\sin(\theta_1)$}
   \KCfor{$n$}{$2,\ldots,N-1$}{
      \KClet{$\theta_n$}{$\arccos(a_n/s_{n-1})$}
      \KClet{$s_n$}{$s_{n-1}\sin(\theta_n)$}
   }
}
\caption{Computation of complex hyperspherical coordinates}
\label{algo1}
\end{KCalgorithm}

\subsection{Control Assumptions}

The following scheme is based on the assumptions that (a) we can
neglect free evolution $H_0=0$; (b) we have local phase control,
i.e., we can implement control operators that introduce a local
phase shift,
\begin{equation}
 \label{eq:Zn}
  Z_n = \Pi_n, \quad n=2,\ldots, N
\end{equation}
where $I_N$ is the identity on $\H$ and $\Pi_n$ is the projector
onto the subspace of $\H$ spanned by the basis state $\ket{n}$; and
(c) we can individually control transitions between adjacent energy
levels, i.e. that we can realize control Hamiltonians of the form
$X_n$ or $Y_n$,
\begin{subequations}
  \begin{align}
  X_n =  (\ket{n+1}\langle{n}|+\ket{n}\langle{n+1}|), \quad n=1,\ldots,N-1.\\
  Y_n = i(\ket{n+1}\langle{n}|-\ket{n}\langle{n+1}|), \quad
   n=1,\ldots,N-1. \label{eq:Yn}
  \end{align}
\end{subequations}
The evolution of the system under any Hamiltonian $H$ is governed by
the Schr\"{o}dinger equation
\begin{equation}
  i\hbar \dot{U}(t) = H U(t), \quad U(0)=I_N,
\end{equation}
and we choose units such that the Planck constant $\hbar=1$.  This shows
that the evolution under the control Hamiltonian $H \in \{L X_n,LY_n,L
Z_n\}$ is given by the one-parameter groups $\exp(-iLt X_n)$, $\exp(-i
Lt Y_n)$ and $\exp(-i Lt Z_n)$, respectively.  The evolution is unitary
as the operators $X_n$, $Z_n$ and $Y_n$ are Hermitian.  In particular,
this means that we can implement the complex rotations
\begin{equation}
 U_n^X(\alpha)=\exp(-i\alpha X_n),\quad
 U_n^Y(\alpha)=\exp(-i\alpha Y_n),\quad
 U_n^Z(\alpha)=\exp(-i\alpha Z_n),
\end{equation}
by applying the control Hamiltonians $LX_n$, $LY_n$ and $LZ_n$
respectively for some time $t=\alpha/L$.  In the following only two
types of control operations $\{X_n,Z_n\}$ or $\{Y_n,Z_n\}$ are
required.

\begin{figure}
\includegraphics[width=\textwidth]{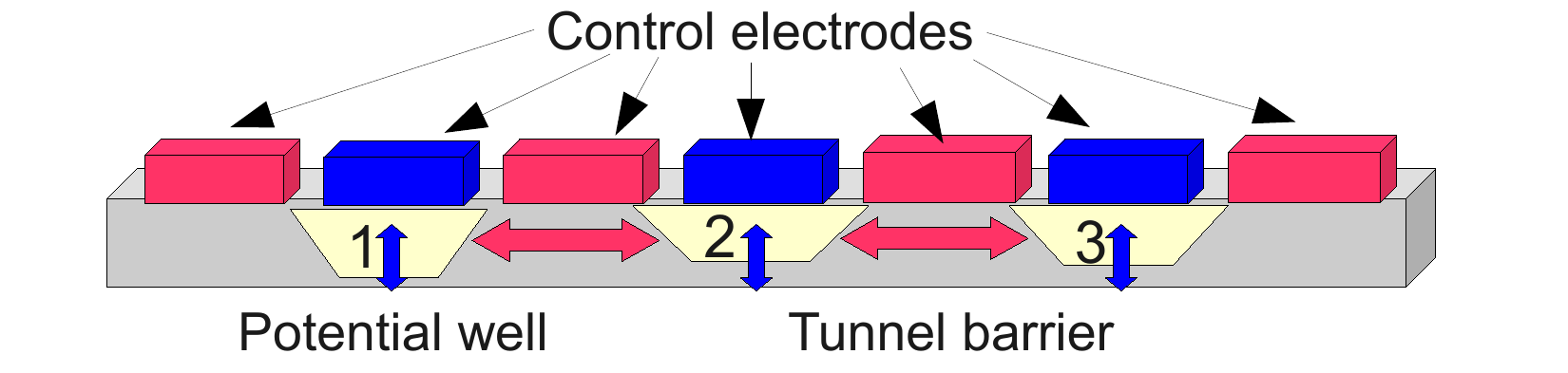}
\caption{Charged particle trapped in a multi-well potential created by
control electrodes.  Red electrodes allow control of potential barriers
and thus tunnelling rates, while blue electrodes control depths of the
wells $1$ to $3$ and thus their energy levels.  We can choose default
voltage settings such that all wells have the same depth and there is no
tunnelling, so that we effectively have $H_0=0$.  Then by raising or
lowering the voltage of electrode $1$ we can introduce a relative phase
shift between the ground state $\ket{1}$ of the first well and the other
two ground states, and by changing the voltage applied to the red
electrode between wells $1$ and $2$ we can induce tunnelling between
the first two wells, etc.}  \label{fig1}
\end{figure}

The assumptions on the control Hamiltonian are somewhat demanding,
although no more so than the control requirements for the standard
geometric decomposition Eq.(\ref{eq:decomp}).  While these requirements
cannot always be satisfied, there are systems for which these control
operations are quite natural such as a charged particle trapped in a
multi-well potential created and controlled by surface control
electrodes as shown in Fig.~\ref{fig1}.  A physical realization of such
a system could be a multi-well potential created in a 2D electron gas in
a semiconductor material by surface control electrodes.  Changing the
voltages applied to different control electrode enables us to vary the
depth of individual wells as well as the height of the potential barrier
between adjacent wells and thus the tunnelling rate, giving raise $Z_n$
and $Y_n$ rotations, respectively.

\subsection{Explicit Control Sequence}

To illustrate the constructive procedure, let us consider the case $N=3$
with $Y,Z$ controls.  In this case the control operators take the
explicit form
\begin{equation}
 Z_2  = \begin{pmatrix} 0&0&0\\ 0&1&0\\ 0&0&0\end{pmatrix}, \quad
 Z_3  = \begin{pmatrix} 0&0&0\\ 0&0&0\\ 0&0&1\end{pmatrix}, \quad
 Y_1  = \begin{pmatrix} 0&-i&0\\i&0&0\\0&0&0\end{pmatrix}, \quad
 Y_2  = \begin{pmatrix} 0&0&0\\ 0&0&-i\\ 0&i&0 \end{pmatrix}. \nonumber
\end{equation}
and the corresponding evolution operators are
\begin{equation}
 U_2^Z(\alpha) = \begin{pmatrix}
          1&0&0\\0&e^{-i\alpha}&0\\0&0&1
         \end{pmatrix}, \quad
 U_3^Z(\alpha) = \begin{pmatrix}
          1&0&0\\0&1&0\\0&0&e^{-i\alpha}
         \end{pmatrix}, \nonumber
\end{equation}

\begin{equation}
 U_1^Y(\alpha) = \begin{pmatrix}
                 \cos\alpha&-\sin\alpha&0\\
                 \sin\alpha&\cos\alpha&0\\
          0&0&1
                 \end{pmatrix}, \quad
 U_2^Y(\alpha) = \begin{pmatrix}
          1&0&0\\
          0&\cos\alpha&-\sin\alpha\\
          0&\sin\alpha&\cos\alpha
          \end{pmatrix}, \nonumber
\end{equation}
  Given these control operators and the hyperspherical coordinate
representation of the initial and target states, it is now very easy to
see how to steer an arbitrary initial state to an arbitrary target state
in the following seven steps:

\noindent\textbf{Step 1.}
$(\theta^{(0)}_{1},\theta^{(0)}_{2};\phi^{(0)}_{1},\phi^{(0)}_{2})
\rightarrow(\theta^{(0)}_{1},\theta^{(0)}_{2};\phi^{(0)}_{1},0)$:
Apply phase rotation $U_3^Z(\phi_2^{(0)})$
\begin{equation} \nonumber
\begin{pmatrix}
1&0&0\\ 0&1&0\\ 0&0&e^{-i\phi_{2}^{(0)}}
\end{pmatrix}
\begin{pmatrix}
\cos\theta_{1}^{(0)}\\
e^{i\phi_{1}^{(0)}}\sin\theta_{1}^{(0)}\cos\theta_{2}^{(0)}\\
e^{i\phi_{2}^{(0)}}\sin\theta_{1}^{(0)}\sin\theta_{2}^{(0)}
\end{pmatrix}
=\begin{pmatrix}
\cos\theta_{1}^{(0)}\\
 e^{i\phi_{1}^{(0)}}\sin\theta_{1}^{(0)}\cos\theta_{2}^{(0)}\\
                  \sin\theta_{1}^{(0)}\sin\theta_{2}^{(0)}
\end{pmatrix}.
\end{equation}

\noindent\textbf{Step 2.}
$(\theta^{(0)}_{1},\theta^{(0)}_{2};\phi^{(0)}_{1},0)
\rightarrow(\theta^{(0)}_{1},\theta^{(0)}_{2};0,0)$:
Apply phase rotation $U_2^Z(\phi_1^{(0)})$
\begin{equation} \nonumber
\left(\begin{array}{ccc}
1&0&0\\
0&e^{-i\phi_{1}^{(0)}}&0\\
0&0&1
\end{array}\right)
\left(\begin{array}{c}
\cos\theta_{1}^{(0)}\\
e^{i\phi_{1}^{(0)}}\sin\theta_{1}^{(0)}\cos\theta_{2}^{(0)}\\
\sin\theta_{1}^{(0)}\sin\theta_{2}^{(0)}
\end{array}\right)=\left(\begin{array}{c}
\cos\theta_{1}^{(0)}\\
\sin\theta_{1}^{(0)}\cos\theta_{2}^{(0)}\\
\sin\theta_{1}^{(0)}\sin\theta_{2}^{(0)}
\end{array}\right)
\end{equation}

\noindent\textbf{Step 3.}
$(\theta^{(0)}_{1},\theta^{(0)}_{2};0,0)\rightarrow(\theta^{(0)}_{1},0;0,0)$:
Apply population rotation $U_2^Y(-\theta_2^{(0)})$
\begin{equation} \nonumber
\left(\begin{array}{ccc}
1&0&0\\
0&\cos\theta_{2}^{(0)}&\sin\theta_{2}^{(0)}\\
0&-\sin\theta_{2}^{(0)}&\cos\theta_{2}^{(0)}
\end{array}\right)
\left(\begin{array}{c}
\cos\theta_{1}^{(0)}\\
\sin\theta_{1}^{(0)}\cos\theta_{2}^{(0)}\\
\sin\theta_{1}^{(0)}\sin\theta_{2}^{(0)}
\end{array}\right)=\left(\begin{array}{c}
\cos\theta_{1}^{(0)}\\
\sin\theta_{1}^{(0)}\\
0
\end{array}\right)
\end{equation}

\noindent\textbf{Step 4.}
$(\theta^{(0)}_{1},0;0,0)\rightarrow(\theta^{(s)}_{1},,0;0,0)$:
Apply population rotation $U_1^Y(\theta_1^{(s)}-\theta_1^{(0)})$
\begin{equation}\nonumber
\left(\begin{array}{ccc}
\cos(\theta_{1}^{(s)}-\theta_{1}^{(0)})&-\sin(\theta_{1}^{(s)}-\theta_{1}^{(0)})&0\\
\sin(\theta_{1}^{(s)}-\theta_{1}^{(0)})&\cos(\theta_{1}^{(s)}-\theta_{1}^{(0)})&0\\
0&0&1
\end{array}\right)
\left(\begin{array}{c}
\cos\theta_{1}^{(0)}\\
\sin\theta_{1}^{(0)}\\
0
\end{array}\right)=\left(\begin{array}{c}
\cos\theta_{1}^{(s)}\\
\sin\theta_{1}^{(s)}\\
0
\end{array}\right)
\end{equation}

\noindent\textbf{Step 5.}
$(\theta^{(s)}_{1},0;0,0)\rightarrow(\theta^{(s)}_{1},\theta^{(s)}_{2};0,0)$:
Apply population rotation $U_2^Y(\theta_2^{(s)})$
\begin{equation}\nonumber
\left(\begin{array}{ccc}
1&0&0\\
0&\cos\theta_{2}^{(s)}&-\sin\theta_{2}^{(s)}\\
0&\sin\theta_{2}^{(s)}&\cos\theta_{2}^{(s)}
\end{array}\right)
\left(\begin{array}{c}
\cos\theta_{1}^{(s)}\\
\sin\theta_{1}^{(s)}\\
0
\end{array}\right)=\left(\begin{array}{c}
\cos\theta_{1}^{(s)}\\
\sin\theta_{1}^{(s)}\cos\theta_{2}^{(s)}\\
\sin\theta_{1}^{(s)}\sin\theta_{2}^{(s)}
\end{array}\right)
\end{equation}

\noindent\textbf{Step 6.}
$(\theta^{(s)}_{1},\theta^{(s)}_{2};0,0)\rightarrow(\theta^{(s)}_{1},\theta^{(s)}_{2};\phi^{(s)}_{1},0)$:
Apply phase rotation $U_2^Z(-\phi_1^{(s)})$
\begin{equation}\nonumber
\left(\begin{array}{ccc}
1&0&0\\
0&e^{i\phi_{1}^{(s)}}&0\\
0&0&1
\end{array}\right)
\left(\begin{array}{c}
\cos\theta_{1}^{(s)}\\
\sin\theta_{1}^{(s)}\cos\theta_{2}^{(s)}\\
\sin\theta_{1}^{(s)}\sin\theta_{2}^{(s)}
\end{array}\right)=\left(\begin{array}{c}
\cos\theta_{1}^{(s)}\\
e^{i\phi_{1}^{(s)}}\sin\theta_{1}^{(s)}\cos\theta_{2}^{(s)}\\
\sin\theta_{1}^{(s)}\sin\theta_{2}^{(s)}
\end{array}\right)
\end{equation}

\noindent\textbf{Step 7.}
$(\theta^{(s)}_{1},\theta^{(s)}_{2};\phi^{(s)}_{1},0)\rightarrow(\theta^{(s)}_{1},\theta^{(s)}_{2};\phi^{(s)}_{1},\phi^{(s)}_{2})$:
Apply phase rotation $U_3^Z(-\phi_2^{(s)})$
\begin{equation}\nonumber
\left(\begin{array}{ccc}
1&0&0\\
0&1&0\\
0&0&e^{i\phi_{2}^{(s)}}
\end{array}\right)
\left(\begin{array}{c}
\cos\theta_{1}^{(s)}\\
e^{i\phi_{1}^{(s)}}\sin\theta_{1}^{(s)}\cos\theta_{2}^{(s)}\\
\sin\theta_{1}^{(s)}\sin\theta_{2}^{(s)}
\end{array}\right)=\left(\begin{array}{c}
\cos\theta_{1}^{(s)}\\
e^{i\phi_{1}^{(s)}}\sin\theta_{1}^{(s)}\cos\theta_{2}^{(s)}\\
e^{i\phi_{2}^{(s)}}\sin\theta_{1}^{(s)}\sin\theta_{2}^{(s)}
\end{array}\right)
\end{equation}

The generalization to $N>3$ is straightforward, as shown in
Algorithm~\ref{algo2}.  Given a Hamiltonian of the form
\begin{equation}
  \label{eq:H}
  H = \sum_{m=1}^{2N-1} u_m(t) H_m
\end{equation}
where $H_{2n-1}=Z_{n+1}$, $H_{2n}=Y_n$ and $u_m(t)$ are controls
(e.g. voltages), the bang-bang control sequence given by
Algorithm~\ref{algo2} can be implemented by applying $4N-5$ control
pulses. At the $k$th step we apply a constant control field
$u_{S(k)}=L_k$ for time $t_k=\gamma_k/L_k$, while all other controls
are set to $0$ (or the voltages are set to their default values).
Notice that in practice we cannot apply fields for negative times,
thus the sign of $L_k$ must match that of $\gamma_k$.  However, if
$\gamma_k$ is negative and $L_k>0$, we can also apply a field
$f_{S(k)}=L_k$ for time $t_k=(\gamma_k+2\pi)/L_k$ as
$\gamma_k+2\pi>0$ and effects the same rotation.

If $X_n$ control Hamiltonians are used instead of $Y_n$ control
Hamiltonians, error $i^{n-1}$ is created in the $n$th coordinate by
the population rotations. So the algorithm need be slightly modified
to correct phase factors of $|n\rangle(n=2,\cdots,N)$.  We can
achieve this by adding $\tfrac{\pi}{2}(n\mod 4)$ to the phase angles
$\phi_n$, noting that $e^{i\pi/2 (n \mod 4)}=i^{n}$ and the phase
factor of the $n$th coordinate is $e^{i\phi_{n-1}}$.

Besides giving explicit expressions for the rotation angles in the
decomposition, the scheme has an additional advantage compared the
the standard decomposition Eq.(\ref{eq:decomp}) considered earlier:
While the rotations in the standard factorization do not commute,
the first $N-1$ and final $N-1$ phase rotations in the decomposition
based on complex hyperspherical coordinates are represented by
diagonal matrices which commute.  This means that these operations
can be applied concurrently rather than sequentially, leading to a
potentially considerable reduction in the total length of the
control sequence.

\begin{KCalgorithm}
\KCin{$c^{(0)},c^{(s)}$}{initial and target state vectors}
\KCout{$S,\gamma$}{Bang-bang control sequence}
\KCcode{StateTransfer}{Compute sequence of
      rotations required for state transfer}{99}{\textwidth}{
   \KClet{$(\theta^{(0)},\phi^{(0)})$}{\KCid{HyperCoord}$(c^{(0)})$}
   \KClet{$(\theta^{(s)},\phi^{(s)})$}{\KCid{HyperCoord}$(c^{(s)})$}
   \KCfor{$n$}{$N-1,\ldots,1$}{
       \KCstate{Append $S$ by $2n-1$, $\gamma$ by $\phi_n^{(0)}$}%
       [Apply Phase Rotation $U_{n+1}^Z(\phi_n^{(0)})$]}
   \KCfor{$n$}{$N-1,\ldots,2$}{
       \KCstate{Append $S$ by $2n$, $\gamma$ by $-\theta_n^{(0)}$}%
       [Apply Population Rotation $U_{n}^Y(-\theta_n^{(0)})$]}
   \KCstate{Append $S$ by $2$, $\gamma$ by $\theta_1^{(s)}-\theta_1^{(0)}$}%
   [Apply Population Rotation $U_{1}^Y(\theta_1^{(s)}-\theta_1^{(0)})$]
   \KCfor{$n$}{$2,\ldots,N-1$}{
       \KCstate{Append $S$ by $2n$, $\gamma$ by $\theta_n^{(s)}$}%
       [Apply Population Rotation $U_{n}^Y(\theta_n^{(s)})$]}
   \KCfor{$n$}{$1,\ldots,N-1$}{
       \KCstate{Append $S$ by $2n-1$, $\gamma$ by $-\phi_n^{(s)}$}%
        [Apply Phase Rotation $U_{n+1}^Z(-\phi_n^{(s)})$]}
}
%%%%%%%%%%%%%%%%%%%%%%%%%%%%%%%%%%%%%%%%%%%%%%%%%%%%%%%%%%%%%%%%
\caption{Control Scheme to achieve state transfer $\vc^{(0)}\mapsto
\vc^{(s)}$ in $4N-5$ steps using bang-bang control, based on
hyperspherical coordinate parametrization. $\vec{S}$ and $\vec{\gamma}$
are vectors of length $4N-5$, whose elements are integer labels
indicating the control Hamiltonian ($m=1,\ldots,2N-2$) and rotation
angle $\gamma_k$, respectively.}\label{algo2}
\end{KCalgorithm}

\subsection{Application: Creating a multi-partite-entangled $W$-state}

As a simple application of the scheme, suppose first that we have $N$
sites and starting with only site $1$ populated, i.e., in state
$\ket{1}$, we would like to prepare an equal superposition of all $N$
sites $\ket{n}$ for $n=1,\ldots,N$:
\begin{equation*}
   \ket{\psi} = \tfrac{1}{\sqrt{N}} \sum_{n=1}^N \ket{n}.
\end{equation*}
All we need to do is compute the hyperspherical coordinates of
$\ket{\psi}$, e.g., for $N=10$
\begin{equation*}
\vec{\theta} =
  (1.2490, 1.2310, 1.2094, 1.1832, 1.1503, 1.1071, 1.0472, 0.9553, 0.7854)
\end{equation*}
and here clearly $\vec{\varphi}=\vec{0}$, which tells us that we need
to apply a sequence of $9$ $Y$-rotations
\begin{equation*}
  U^Y_{9}(\theta_9)U_8^Y(\theta_8)U_7^Y(\theta_7)U_6^Y(\theta_6)
  U_5^Y(\theta_5) U_4^Y(\theta_4) U_3^Y(\theta_3) U_2^Y(\theta_2) U_1^Y(\theta_1)
\end{equation*}
to the initial state $\ket{1}$.  This results in the following sequence
of states being created
\begin{center}
\footnotesize
\begin{tabular}{*{10}{|r}|}
\hline
    $n=0$  & $n=1$  & $n=2$  & $n=3$  & $n=4$  & $n=5$  & $n=6$  & $n=7$  & $n=8$  & $n=9$ \\\hline
    1.0000 & 0.3162 & 0.3162 & 0.3162 & 0.3162 & 0.3162 & 0.3162 & 0.3162 & 0.3162 & 0.3162\\
         0 & 0.9487 & 0.3162 & 0.3162 & 0.3162 & 0.3162 & 0.3162 & 0.3162 & 0.3162 & 0.3162\\
         0 &      0 & 0.8944 & 0.3162 & 0.3162 & 0.3162 & 0.3162 & 0.3162 & 0.3162 & 0.3162\\
         0 &      0 &      0 & 0.8367 & 0.3162 & 0.3162 & 0.3162 & 0.3162 & 0.3162 & 0.3162\\
         0 &      0 &      0 &      0 & 0.7746 & 0.3162 & 0.3162 & 0.3162 & 0.3162 & 0.3162\\
         0 &      0 &      0 &      0 &      0 & 0.7071 & 0.3162 & 0.3162 & 0.3162 & 0.3162\\
         0 &      0 &      0 &      0 &      0 &      0 & 0.6325 & 0.3162 & 0.3162 & 0.3162\\
         0 &      0 &      0 &      0 &      0 &      0 &      0 & 0.5477 & 0.3162 & 0.3162\\
         0 &      0 &      0 &      0 &      0 &      0 &      0 &      0 & 0.4472 & 0.3162\\
         0 &      0 &      0 &      0 &      0 &      0 &      0 &      0 &      0 & 0.3162\\\hline
\end{tabular}
\end{center}
Each $Y_n$ extends the superposition by one site until we are left with
the desired state after the final step.  Creating such a superposition
state may not seem very interesting but it has an interesting
application in the area of entanglement creation, for instance.

The Hamiltonians for many systems such as interacting quantum dots or
coupled Josephson junctions, for example, can be described to a
reasonably good approximation by an XXZ-spin network model
\begin{equation}
  \label{eq:HXXZ}
  H = \sum_{n} \alpha_n \sigma^Z_n + \sum_{m<n}\gamma_{mn} (\sigma^X_m
  \sigma^X_n + \sigma^Y_m \sigma^Y_n + \kappa \sigma^Z_m \sigma^Z_n),
\end{equation}
where $\sigma^A_n$ is an $N$-fold tensor product for which the $n$th
factor is $A$ and all others are the identity, and $X$, $Y$ and $Z$ are
the usual $2\times 2$ Pauli matrices.  For a chain with nearest-coupling
we have $\gamma_{mn}=0$ except when $n=m+1$ and for $\kappa=0$ we have
the so-called XX-coupling model.  The Hamiltonian (\ref{eq:HXXZ}) commutes
with the total spin operator $S=\sum_n Z_n$ and decomposes into
excitation subspaces for any choice of $\alpha_n$ and $\gamma_{mn}$.  It
is easy to see that a state such as the $W$-state
\begin{equation}
   \psi_W = \tfrac{1}{\sqrt{N}} \left(\ket{\up\dn\dn\ldots\dn} +
    \ket{\dn\up\dn\ldots \dn} + \ldots + \ket{\dn\dn\ldots \dn\up} \right)
\end{equation}
belongs to the single excitation subspace, as does the state
$\ket{\up\dn\ldots\dn}$.  On this subspace the Hamiltonian (\ref{eq:HXXZ})
can be simplified.  For a chain with nearest-neighbour coupling and
$\kappa=0$ we obtain (up to multiples of the identity):
\begin{equation}
  \label{eq:H1}
  H_1 = \sum_{n} -\alpha_n Z_n + \gamma_{n} X_n,
\end{equation}
i.e., the Hamiltonian is exactly of the form required for our scheme.
Treating $\alpha_n$ and $\gamma_n$ as control parameters, we can use
the scheme above to create a $W$ state starting from the product state
$\ket{1}=\ket{\up\dn\ldots\dn}$.  Since we can only implement
$X_n$-rotations, we will need to apply the phase corrections
\begin{equation*}
  \prod_{n=2}^N U_n^Z(\phi_n), \quad \phi_n = -\tfrac{\pi}{2} (n-1 \mod 4),
\end{equation*}
which can be applied concurrently in the final step.

\section{Optimal piecewise-constant Control and Time-energy Performance}

The bang-bang control sequence given by Algorithm~\ref{algo2} leaves
us considerable freedom of choice for the controls.  Choosing large
control amplitudes will result in short pulse durations, thus
optimizing the transfer time $t_f$. However, large control
amplitudes may not be feasible and have undesirable side effects in
terms of transfering too much energy to the system.  We can try to
optimize the field amplitude by stipulating that the state transfer
is to be achieved while minimizing a time-energy performance index
\begin{equation}
  \label{2-16}
   J  = \int_{0}^{t_{f}}[\lambda+\sum_{m=1}^{2N-2} |u_m(t)|^2] \, dt
\end{equation}
where $\lambda$ is the ratio factor of the costs of time and energy
and $\lambda>0$. Larger values of $\lambda$ indicate a stronger
emphasis on time-cost, while smaller values of $\lambda$ give more
weight to the energy cost of the controls.

If the controls can take values $f_m(t)\in\{0,\pm L\}$ and the
pulses are applied strictly sequentially, then the total length
$t_f$ of the control sequence is
\begin{equation}
 \begin{split}
  t_f &= \frac{1}{L} \left[
         \sum_{n=1}^{N-1} |\phi_n^{(0)}| + |\phi_n^{(s)}|
         + \sum_{n=2}^{N-1} (\theta_n^{(0)} + \theta_n^{(s)})
         + |\theta_1^{(s)}- \theta_1^{(0)}|\right]\\
      & \le \frac{1}{L}\left[2(N-1)\pi + 2(N-2)\frac{\pi}{2} +
            \frac{\pi}{2}\right]
        = \frac{(6N-7)\pi}{2L}
\end{split}
\end{equation}
because of $0\le\theta_n\le\tfrac{\pi}{2}$ and $0\le|\phi_n|\le
\pi$. Noting that $a^2+b^2 \ge 2ab$, with equality exactly if $a=b$,
we have
\begin{equation}
 \label{Jb}
 J = \sum_{k=1}^K (\lambda + L_k^2)t_k
   \le \sum_{k=1}^K 2\sqrt{\lambda} L_k t_k
   \le 2 \sqrt{\lambda} t_f \max_k L_k
\end{equation}
with equality if and only if $L_k=\sqrt{\lambda}$.  This shows that the
optimal choice of the field amplitudes is $L_k=\sqrt{\lambda}$, for
which we have
\begin{equation}
 t_f^*\le\frac{(6N-7)\pi}{2\sqrt{\lambda}}, \quad
 J_* = \min J = 2\lambda t_f^* \le \sqrt{\lambda} (6N-7)\pi
 \end{equation}
and the corresponding optimal energy cost is $E^*=J^*-\lambda t_f^*\le
\frac{1}{2}\sqrt{\lambda} (6N-7)\pi$. As expected, as $\lambda$ goes to
$0$, $t_f^*$ becomes infinite and $E^*$ goes to $0$, but their product
remains constant
\begin{equation}
\label{constant} \begin{array}{c}
  t_{f}^{*}\cdot{E^{*}}=
  \left[ \sum_{n=1}^{N-1} |\phi_n^{(0)}| + |\phi_n^{(s)}|
         + \sum_{n=2}^{N-1} (\theta_n^{(0)} + \theta_n^{(s)})
         + |\theta_1^{(s)}- \theta_1^{(0)}|\right]^2
  \leq \frac{(6N-7)^2\pi^2}{4}
\end{array}
\end{equation}
and depends only on the geometric parameters of the initial state and
target states.

If first and last $N-1$ phase rotations are applied concurrently the
transfer time is reduced
\begin{equation}
 \begin{split}
  t_f' &= \frac{1}{L} \left[
         \max_n |\phi_n^{(0)}| + \max_n |\phi_n^{(s)}|
         + \sum_{n=2}^{N-1} (\theta_n^{(0)} + \theta_n^{(s)})
         + |\theta_1^{(s)}- \theta_1^{(0)}|\right]\\
      & \le \frac{1}{L}\left[2\pi + 2(N-2)\frac{\pi}{2} +
            \frac{\pi}{2}\right]
        = \frac{(2N+3)\pi}{2L}.
\end{split}
\end{equation}
Setting $\phi_{\max}^{(0)}=\max_n |\phi_n^{(0)}|$ and
$\phi_{\max}^{(s)}=\max_n |\phi_n^{(s)}|$, shows that we have $t_1 =
\phi_{\max}^{(0)}/L$ and $t_K=\phi_{\max}^{(s)}/L$, and thus we must
choose $L_n\ge \phi_n^{(0)}/t_1$ and $L_n=\phi_n^{(s)}/t_K$,
respectively for the control amplitude of the first and last $N-1$
concurrent pulses to be able to implement all $N-1$ phase rotations
concurrently in time $t_1$ or $t_K$, respectively.  Furthermore the
performance index changes
\begin{equation}
  J \le 2 t_f' \sqrt{\lambda} \max_{N\le k\le K+1-N} L_k
    + \sum_{k=1}^{N-1} L_k^2 + \sum_{k=K-N+2}^K L_k^2,
\end{equation}
which suggests that we can improve the performance index and reduce
the energy cost by choosing the amplitudes of the first and last
$N-1$ concurrent pulses to be as small as possible, i.e.,
$L_n=\phi_n^{(0)}/t_1$ and $L_n=\phi_n^{(s)}/t_K$, and
$L_k=\sqrt{\lambda}$ for all other amplitudes.

\section{Implementation of Unitary Operators}

\subsection{Complex hyperspherical representation of unitary operators}

Any $N-$dimensional unitary operator can be represented as follows:
\begin{equation}
\label{ru-1} U=\sum_{j=1}^N e^{i\varphi_j} \ket{u_j} \bra{u_j}
\end{equation}
where $\{\ket{u_j}\}$ constructs an orthonormal basis set in
$N-$dimensional Hilbert space. From~(\ref{ru-1}), we can find that
global phase factors of all $\ket{u_j}$s do not affect $U$, so
they can be neglected.  Assuming $\ip{1}{u_j}$ is real
and positive for each $j$, by~(\ref{3-2}) the complex hyperspherical
parametrization for $\{\ket{u_j}: j=1,\cdots,N\}$ can be given by
\begin{equation}
\label{ru-2}
 \begin{pmatrix}
   \ket{u_1}\\ \ket{u_2} \\ \vdots\\ \ket{u_N}
 \end{pmatrix}
  =  \begin{pmatrix} \mathbf{I}_{N-2} & 0 \\ 0 & C^{(2)} \end{pmatrix}
     \cdots
     \begin{pmatrix} \mathbf{I}_{1} & 0 \\ 0 & C^{(N-1)} \end{pmatrix}
      C^{(N)}
      \begin{pmatrix}
      \ket{1} \\ \ket{2} \\ \vdots\\ \ket{N}
      \end{pmatrix}
\end{equation}
where
\begin{equation}
\label{ru-3}
         C^{(k)} = \left(
                   \overrightarrow{c_1}^{(k)}),
           (\overrightarrow{c_2}^{(k)}, \ldots,
           (\overrightarrow{c_{k-1}}^{(k)}),
           (\overrightarrow{c_k}^{(k)}) \right)^T,
\end{equation}
and $(\overrightarrow{c_i}^{(k)})^T$ is the transpose of the vector
$\overrightarrow{c_i}^{(k)}$ and
\begin{align}
\label{ru-4.1}
 \overrightarrow{c_1}^{(k)}
 &= \begin{pmatrix}
     \cos\theta_1^{(k)}\\
     e^{i\phi_1^{(k)}}\sin\theta_1^{(k)}\cos\theta_2^{(k)}\\
     \vdots\\
     e^{i\phi_{k-2}^{(k)}}\prod_{l=1}^{k-2}\sin\theta_l^{(k)}\cos\theta_{k-1}^{(k)}\\
     e^{i\phi_{k-1}^{(k)}}\prod_{l=1}^{k-1}\sin\theta_l^{(k)}
    \end{pmatrix},
\end{align}
\begin{align}
\overrightarrow{c_2}^{(k)}
 &= \begin{pmatrix}
     \sin\theta_1^{(k)}\\
     -e^{i\phi_1^{(k)}}\cos\theta_1^{(k)}\cos\theta_2^{(k)}\\
     \vdots\\
     -e^{i\phi_{k-2}^{(k)}}\cos\theta_1^{(k)}\prod_{l=2}^{k-2}\sin\theta_l^{(k)}\cos\theta_{k-1}^{(k)}\\
      -e^{i\phi_{k-1}^{(k)}}\cos\theta_1^{(k)}\prod_{l=2}^{k-1}\sin\theta_l^{(k)}
    \end{pmatrix},
\end{align}
and so forth until
\begin{align}
\overrightarrow{c_{k-1}}^{(k)}
 &= \begin{pmatrix}
     0\\
     \vdots\\
     0\\
     e^{i\phi_{k-3}^{(k)}}\sin\theta_{k-2}^{(k)}\\
     -e^{i\phi_{k-2}^{(k)}}\cos\theta_{k-2}^{(k)}\cos\theta_{k-1}^{(k)}\\
     -e^{i\phi_{k-1}^{(k)}}\cos\theta_{k-2}^{(k)}\sin\theta_{k-1}^{(k)}
    \end{pmatrix}, \label{ru-4.2}
\end{align}
\begin{align}
 \overrightarrow{c_k}^{(k)}
 &= \begin{pmatrix}
     0\\
     \vdots\\
     0\\
     e^{i\phi_{k-2}^{(k)}}\sin\theta_{k-1}^{(k)}\\
    -e^{i\phi_{k-1}^{(k)}}\cos\theta_{k-1}^{(k)}
    \end{pmatrix}
\end{align}
for $2\leq k\leq N$.  Note that $C^{(k)}(C^{(k)})^\dag=\mathbf{I}_k$.

\subsection{Realization of unitary operators}

Suppose that $Y_n$ and $Z_n$ controls as defined in eqs.~(\ref{eq:Zn})
and (\ref{eq:Yn}) are permitted for $1\leq n\leq N-1$ and $1\leq n\leq
N$, respectively.  Then $U$ can be realized by a sequence of bang-bang
controls
\begin{equation}
\label{ru-6}
 U = T^\dag \underline{\left(\prod_{n=1}^{N} U_n^Z(\varphi_n)\right)}T
\end{equation}
where
\begin{equation}
\label{ru-6.1} T=\prod_{n=2}^{N}U_n, \quad
  U_n =  \prod_{j=1}^{n-1}U_{N-n+j}^Y(\theta_j^{(n)})
         \underline{\prod_{j=1}^{n-1}U_{N-n+j+1}^Z(\phi_j^{(n)})}.
\end{equation}
The $Z$-phase rotations (underlined) commute and can be applied
concurrently.  Thus $U_n$ can be implemented in $n$ steps and $T$ in
$N(N+1)/2-1$ steps and the entire process in $N(N+1)-1$ steps.  If we
took the more contentional approach of factoring an operator
$U\in \SU(N)$ into a sequence of $N(N+1)/2$ rotations on two-level
subspaces, e.g., spanned by $\{\ket{n},\ket{n+1}\}$, and further
decomposed each of these $\SU(2)$ rotations into three elementary
$Y_n$ and $Z_n$ rotations using the Euler decomposition, we would
require $3N(N+1)/2$ steps instead, and since $Z_n$ and $Y_n$ operations
do not commute, these could not be implemented concurrently.

The proof of the result is constructive.

\noindent\textbf{(1) Effect of each $\mathbf{U_n}$.}  Let
$\ket{e^{(n)}}$ be an arbitrary state in the space spanned by
$\{\ket{1},\ket{2},\cdots,\ket{N-n}\}$ and
\begin{equation*}
\begin{pmatrix}
\ket{e_{N-n+1}^{(n)}}\\
\ket{e_{N-n+2}^{(n)}}\\ \vdots\\
\ket{e_N^{(n)}}
\end{pmatrix}
= C^{(n)}
\begin{pmatrix}
\ket{N-n+1}\\ \ket{N-n+2}\\ \vdots\\ \ket{N}
\end{pmatrix}
\end{equation*}
where $C^{(n)}$ is as defined in Eq.(\ref{ru-3}). That is,
\begin{align*}
\ket{e_{N-n+1}^{(n)}}
 =& \cos\theta_1^{(n)} \ket{N-n+1}
    + e^{i\phi_1^{(n)}}\sin\theta_1^{(n)}\cos\theta_2^{(n)} \ket{N-n+2}\\
  & + \cdots + e^{i\phi_{n-2}^{(n)}}\sin\theta_{1}^{(n)} \cdots
       \sin\theta_{n-2}^{(n)}\cos\theta_{n-1}^{(n)} \ket{N-1}\\
  & + e^{i\phi_{n-1}^{(n)}}\sin\theta_{1}^{(n)}\cdots\sin\theta_{n-2}^{(n)}
      \sin\theta_{n-1}^{(n)} \ket{N} \\
\ket{e_{N-n+2}^{(n)}}
 =& \sin\theta_1^{(n)}\ket{N-n+1}-e^{i\phi_1^{(n)}}\cos\theta_1^{(n)}\cos\theta_2^{(n)}\ket{N-n+2}\\
  & -\cdots-e^{i\phi_{n-2}^{(n)}}\cos\theta_{1}^{(n)}\sin\theta_{2}^{(n)}
      \cdots\sin\theta_{n-2}^{(n)}\cos\theta_{n-1}^{(n)}\ket{N-1} \\
  & -e^{i\phi_{n-1}^{(n)}}\cos\theta_{1}^{(n)}\sin\theta_2^{(n)}\cdots
     \sin\theta_{n-2}^{(n)}\sin\theta_{n-1}^{(n)}\ket{N}\\
  & \cdots\cdots \\
\ket{e_{N-1}^{(n)}}
 =& e^{i\phi_{n-3}^{(n)}}\sin\theta_{n-2}^{(n)} \ket{N-2} \\
  &-e^{i\phi_{n-2}^{(n)}}\cos\theta_{n-2}^{(n)}\cos\theta_{n-1}^{(n)}\ket{N-1}
   -e^{i\phi_{n-1}^{(n)}}\cos\theta_{n-2}^{(n)}\sin\theta_{n-1}^{(n)}\ket{N}\\
\ket{e_N^{(n)}}
 =& e^{i\phi_{n-2}^{(n)}}\sin\theta_{n-1}^{(n)}\ket{N-1}
   -e^{i\phi_{n-1}^{(n)}}\cos\theta_{n-1}^{(n)}\ket{N}.
\end{align*}

$U_n$ leaves any state $\ket{e^{(n)}}$ in the subspace spanned by
$\{\ket{1},\ket{2},\cdots,\ket{N-n}\}$ invariant, i.e., $U_n
\ket{e^{(n)}}=\ket{e^{(n)}}$ as $U_n$ is the identity on this subspace.
Furthermore, Sec.~3.3 shows that applying $U_n$ to $\ket{e_{N-n+1}}$
maps it to the basis state $\ket{N-n+1}$, and we can verify by direct
computation
\begin{align*}
  U_n\ket{e_{N-n+2}^{(n)}}
 &= U_{N-n+1}^Y(\theta_{1}^{(n)})(\sin\theta_{1}^{(n)}\ket{N-n+1}
    -\cos\theta_1^{(n)}\ket{N-n+2})\\
 &= -\ket{N-n+2} \\
 & \cdots\cdots \\
 U_n\ket{e_{N-1}^{(n)}}
 &= \prod_{j=1}^{n-2}U_{N-n+j}^Y(\theta_j^{(n)})(\sin\theta_{n-2}^{(n)}\ket{N-2}
    -\cos\theta_{n-2}^{(n)}\ket{N-1})\\
 &= -\ket{N-1}\\
U_n \ket{e_{N}^{(n)}}
 &= \prod_{j=1}^{n-1}U_{N-n+j}^Y(\theta_j^{(n)})(\sin\theta_{n-1}^{(n)}
\ket{N-1}-\cos\theta_{n-1}^{(n)}\ket{N-2})\\
 &= -\ket{N}
\end{align*}
\begin{equation}
\label{ru-8}
\begin{cases}  \begin{array}{ll}
 U_n \ket{e^{(n)}}= \ket{e^{(n)}}
 & \ket{e^{(n)}}\in\mbox{\rm Span}\{\ket{1}, \cdots,\ket{N-n}\}\\
 U_n \ket{e_{N-n+1}^{(n)}} =  \ket{N-n+1} \\
 U_n \ket{e_{j}^{(n)}}     = -\ket{j}
 & N-n+2\leq j\leq N.
\end{array}\end{cases}
\end{equation}

\noindent\textbf{(2) Effect of $\mathbf{T}$.}  Using the previous result
we now show that $T\ket{u_n}=(-1)^{(n-1)}\ket{n}$ with $\ket{u_n}$ as
defined in (\ref{ru-2}).  Let $\vec{a}^{(k)}=(0,\cdots,0,
(\overrightarrow{c_1}^{(k)})^T)$ be a row vector of length $N$ where the
coefficient vector $c_1^{(k)}$ is as in Eq.~(\ref{ru-4.1}) and the
number of zeros is $N-k$.  Eqs~(\ref{ru-2})--(\ref{ru-6.1}) and
(\ref{ru-8}) give
\begin{align*}
T \ket{u_1} &= U_1\cdots U_N \vec{a}^{(N)}
               \begin{pmatrix}\ket{1}\\ \vdots\\\ket{N} \end{pmatrix}\\
            &= U_1\cdots U_{N-1}U_N \ket{e_1^{(N)}}
             = U_1\cdots U_{N-1}\ket{1}=\ket{1}
\end{align*}
\begin{align*}
T \ket{u_2} &= U_1\cdots U_N \vec{a}^{(N-1)}C^{(N)}
               \begin{pmatrix} \ket{1} \\ \vdots\\ \ket{N} \end{pmatrix}
             = U_1\cdots U_{N}\vec{a}^{(N-1)}
               \begin{pmatrix}
                \ket{e_1^{(N)}}\\ \vdots\\ \ket{e_N^{(N)}}
               \end{pmatrix} \\
            &= U_1\cdots U_{N-1}\vec{a}^{(N-1)}
               \begin{pmatrix}
        \ket{1} \\-\ket{2} \\ \vdots\\ -\ket{N}
               \end{pmatrix}
             = -U_1\cdots U_{N-1} \ket{e_2^{(N-1)}} =-\ket{2}
\end{align*}
Furthermore, for $3\leq n\leq N-1$ we have
\begin{align*}
U_N \ket{u_n}
%=&U_N\vec{a}^{(N-n+1)}\prod_{k=N-n+2}^{N-1}
%\begin{pmatrix} \mathbf{I}_{N-k}&\\ & C^{(k)}\end{pmatrix} C^{(N)}
%\begin{pmatrix}\ket{1}\\ \vdots\\\ket{N} \end{pmatrix}\\
=&\vec{a}^{(N-n+1)} \prod_{k=N-n+2}^{N-1}
  \begin{pmatrix}\mathbf{I}_{N-k}&\\ & C^{(k)} \end{pmatrix}
  \begin{pmatrix} \ket{1} \\ -\ket{2} \\ \vdots \\ -\ket{N} \end{pmatrix}\\
=&\vec{a}^{(N-n+1)} \prod_{k=N-n+2}^{N-2}
    \begin{pmatrix}\mathbf{I}_{N-k}&\\ & C^{(k)} \end{pmatrix}
    \begin{pmatrix} \ket{1} \\-\ket{e_2^{(N-1)}}\\
    \vdots\\ -\ket{e_N^{(N-1)}} \end{pmatrix} \\
U_{N-1}U_N \ket{u_n}
=& \vec{a}^{(N-n+1)}\prod_{k=N-n+2}^{N-3}
\begin{pmatrix} \mathbf{I}_{N-k}&\\ & C^{(k)}\end{pmatrix}
\begin{pmatrix} \ket{1} \\-\ket{2}\\
(-1)^2\ket{e_3^{(N-2)}}\\ \vdots\\ (-1)^2 \ket{e_N^{(N-2)}}
\end{pmatrix}
\end{align*}
and continuing we obtain for $3\le n \le N-1$
%\begin{equation}
%U_{N-n+4}\cdots U_N|u_n\rangle=\vec{a}^{(N-n+1)}\left(\begin{array}{cc}\mathbf{I}_{n-2}&\\
%&
%C^{(N-n+2)}\end{array}\right)\left(\begin{array}{c}(-1)^0|1\rangle\\
%\vdots\\ (-1)^{(n-4)}|n-3\rangle\\ (-1)^{(n-3)}|e_{n-2}^{N-n+3}\rangle\\
%\vdots\\(-1)^{(n-3)}|e_{N}^{(N-n+3)}\rangle
%\end{array}\right)
%\end{equation}
\begin{align*}
%\label{ru-9}
 U_{N-n+3}\cdots U_N \ket{u_n}
 =& \vec{a}^{(N-n+1)}
\begin{pmatrix}
 (-1)^0\ket{1}\\
 \vdots\\
 (-1)^{(n-3)}\ket{n-2}\\ (-1)^{(n-2)}\ket{e_{n-1}^{(N-n+2)}}\\
 \vdots\\
 (-1)^{(n-2)}\ket{e_{N}^{(N-n+2)}}
\end{pmatrix} \\
%\label{ru-10}
U_{N-n+2}\cdots U_N \ket{u_n}
%=\vec{a}^{(N-n+1)}\left(\begin{array}{c}(-1)^0|1\rangle\\
%\vdots\\ (-1)^{(n-2)}|n-1\rangle\\ (-1)^{(n-1)}|n\rangle\\
%\vdots\\
%(-1)^{(n-1)}|N\rangle\end{array}\right)
=& (-1)^{(n-1)}\ket{e_n^{(N-n+1)}} \\
T \ket{u_n} = U_{N-n+1}\cdots U_N \ket{u_n}
=&(-1)^{(n-1)} \ket{n},
\end{align*}
Finally, we have
\begin{equation*}
 U_{3}\cdots U_N\prod_{k=3}^{N-2}
 \begin{pmatrix}\mathbf{I}_{N-k}&\\&C^{(k)} \end{pmatrix} C^{(N)}
 \begin{pmatrix} \ket{1} \\ \ket{2} \\ \vdots \\ \ket{N} \end{pmatrix}
 = \begin{pmatrix}
   (-1)^0 \ket{1} \\ \vdots\\ %(-1)^{(N-3)}\ket{N-2}\\
   (-1)^{(N-2)}\ket{N-1\rangle}\\(-1)^{(N-2)}\ket{N}
   \end{pmatrix}
\end{equation*}
and thus
\begin{align*}
T\ket{u_N}
=& U_2(0,\cdots, 0, (\overrightarrow{c_2}^{(2)})^T)
   \begin{pmatrix}
   (-1)^0\ket{1}\\ \vdots\\ %(-1)^{(N-3)}\ket{N-2}\\
   (-1)^{(N-2)}\ket{N-1}\\(-1)^{(N-2)}\ket{N}
   \end{pmatrix} \\
=& U_2(-1)^{(N-2)} \ket{e_N^{(2)}} = (-1)^{(N-1)}\ket{N}.
\end{align*}
Thus we finally have
\begin{align*}
  T U T^\dag
 = \sum_{n=1}^N e^{i\varphi_n} T \ket{u_n} \bra{u_n} T^\dag
 = \sum_{n=1}^N e^{i\varphi_n} \ket{n}\bra{n}
 = \prod_{n=1}^N U_n^Z(\varphi_n)
\end{align*}
and $U = T^\dag \prod_{n=1}^N U_n^Z(\varphi_n) T$ as claimed.

\section{Discussions and Conclusion}

We have presented an explicit geometric control scheme for quantum
state transfer problems based on a parametrization of the pure state
vectors in terms of complex hyperspherical coordinates.  Although it
is not difficult to find constructive control schemes for state
transfer based on Lie group decompositions, most schemes do not give
explicit expressions for the rotation angles (``generalized Euler
angles'') in the factorization, and thus the rotation angles usually
have to computed numerically.  By parametrizing the initial and 
target states in terms of hyperspherical coordinates, we obtain a
factorization where all generalized Euler angles are given
explicitly in terms of the hyperspherical coordinates of the initial
and target states, eliminating the need for numerical calculation of
the generalized Euler angles, aside from computation of the
hyperspherical coordinates, which is trivial in terms of
computational overhead.

The factorization is applicable given controls capable of implementing
phase rotations and population rotations (of either $X$ or $Y$ type) on
a collection of two-dimensional subspaces, similar to the general
requirements for constructive geometric control schemes.  Compared to
control schemes based on the standard factorization, this scheme has the
additional advantages that all initial and final phase rotations can be
combined in a single step and executed concurrently, reducing the time
required to achieve the state transfer.  As with all bang-bang control
schemes based on Lie group decompositions, the factorization only
determines the sequence in which the controls are applied and the pulse
area (rotation angle) of the control pulses, leaving us with
considerable freedom to choose the pulse shapes and amplitudes, which
can be used to further optimize a performance index.  Here we have
considered optimization of the pulse amplitudes for piecewise constant
controls such as to minimize a time-energy performance index that takes
into account the competing goals of trying to minimize the transfer time
and energy cost of the controls.

The scheme can be generalized to realize unitary operators.  By
expressing the eigenvectors of the target gate $U$ in hyperspherical
coordianates we obtain an explicit decomposition for arbitrary unitary
operators.  Aside from giving explicit expressions for the Euler angles
in terms of hyperspherical coordinates an advantage of the decomposition
is that it separates the elementary rotations in such a way as to allow
concurrent implementation of subsets of operations, which can reduce the
control time.  Specifially, unlike in many standard decomposition
schemes, the $Z_n$ and $Y_n$ rotations do not occur in an alternating
sequence but are clustered.  Since the $Z_n$ rotations are mutually
commuting, this allows concurrent implementation of many operations.

\ack This work is support by National Natural Science Foundationof China
(Grant No. 60974037 \& Grant No. 11074307). SGS acknowledges funding
from EPSRC ARF Grant EP/D07192X/1 and Hitachi.

\end{document}